\documentclass[aps,prl,reprint,superscriptaddress,nofootinbib,floatfix,nobalancelastpage]{revtex4-2}

\usepackage{amsmath,amssymb,amsthm,mathtools,bm,booktabs}
\usepackage{graphicx,tikz,comment}
\usetikzlibrary{arrows.meta,calc,decorations.pathreplacing,positioning} 
\usepackage[colorlinks=true,citecolor=blue,linkcolor=blue,urlcolor=blue]{hyperref}

\usepackage{pgfplots}  \pgfplotsset{compat=1.18}

\newtheorem{theorem}{Theorem}
\newtheorem{corollary}{Corollary}
\newcommand{\C}{\mathbb C}
\newcommand{\ket}[1]{\lvert #1\rangle}
\renewcommand\vec{\boldsymbol}
\newcommand{\NP}[1]{{\color{blue}NP: #1}}

\begin{document}

\title{Entangling power of neural networks}

\author{Taige Wang}
\email{tgwang@mit.edu}
\affiliation{Department of Physics, Massachusetts Institute of Technology, Cambridge, MA-02139, USA}
\affiliation{Department of Physics, Harvard University, Cambridge, MA 02138, USA}

\author{Nisarga Paul}
\email{npaul@caltech.edu}
\affiliation{Department of Physics and Institute for Quantum Information and Matter,
Caltech, Pasadena, CA 91125, USA}

\author{Liang Fu} 
\email{liangfu@mit.edu}
\affiliation{Department of Physics, Massachusetts Institute of Technology, Cambridge, MA-02139, USA}

\begin{abstract}
Characterizing the complexity of correlations between subsystems is a fundamental task across information theory, machine learning, and science. In quantum physics, neural networks have found increasing application in learning wavefunctions. Here we introduce the entangling power of an encoder-decoder neural network, which quantifies its ability to generate entanglement between subsystems, dependent on a latent space dimension $K$ and the complexity class of the decoder. We exactly calculate this quantity for polynomial decoders of degree $p$ acting on a $K$-dimensional latent space. Our results establish the exponential entangling power of neural networks with modest resources.  More broadly, our work provides a framework for analyzing correlations in machine learning that generalizes the notion of the Schmidt rank in entanglement theory. 
\end{abstract}

\maketitle


\paragraph*{\textbf{Introduction.} } Quantifying the complexity of correlations between two sets of variables, in particular across a cut $A|B$, is a common task in classical information theory, quantum information, and quantum many-body physics. In many-body physics, the cut may separate two spatial regions or two groups of particles. In information theory and modern machine learning, a cut may separate two parts of a sequence, two regions of an image, or two blocks of a text corpus~\cite{shannon1948mathematical}. In each setting, the key question is: if the two sides are encoded separately, how much capacity is needed to reconstruct the whole?

For pure quantum states, this question has a canonical answer provided by entanglement theory. A quantum state 
admits a Schmidt decomposition, and the number of (significant) Schmidt coefficients 
defines the Schmidt rank across the cut. This Schmidt rank underlies efficient representations of quantum states using matrix-product states and tensor networks~\cite{eckart1936approximation, ostlund1995}. For instance, many physically relevant ground states exhibit area-law entanglement and can often be represented with modest bond dimension, whereas volume-law states require exponentially many Schmidt coefficients ~\cite{Vidal2003,Hastings2007,Schollwock2011,Orus2014,Cirac2021}.

In recent years, neural networks have emerged as a powerful method for representing quantum wavefunctions $\psi(x)$, where $x=(x_1,\ldots,x_n)$ may be a bitstring representing spin configuration on a lattice, or a list of particle coordinates. A sufficiently large neural network can approximate any continuous function on a compact domain~\cite{hornik1989multilayer,cybenko1989approximation}. Universal neural architectures have also been developed for permutation-symmetric and antisymmetric wavefunctions of bosons and fermions, respectively~\cite{zaheer2017deep, chen2026exact, fu2025minimal, FuFermiSets2026}. However, the ability of neural networks of fixed size to capture quantum wavefunctions remains under active investigation~\cite{deng2017quantum,gao2017efficient,carleo2018constructing,chen2018equivalence,levine2019quantum,huang2021neural,Szabo2020,Yang2024,kufel2025approximately,sinibaldi2025nonstabilizerness,Sharir2022,paul2026bound,WangWalsh2026,lu2026information}.

\begin{figure}[t]
\centering
\pgfplotsset{
  colormap={org}{rgb255=(255,252,247) rgb255=(255,235,200) rgb255=(255,196,110)
                 rgb255=(240,140,40) rgb255=(180,80,20) rgb255=(105,40,15)},
  heat/.style={
    scale only axis, width=3.3cm, height=3.92cm,
    axis lines=left,
    axis line style={thin,-{Stealth[length=1.6mm]}},
    tick style={thin,black!60},
    ticklabel style={font=\scriptsize,color=black!70},
    label style={font=\small,color=black!70},
    every axis x label/.style={at={(ticklabel* cs:1.0)},anchor=west,xshift=2pt},
    every axis y label/.style={at={(ticklabel* cs:1.0)},anchor=south,yshift=2pt},
  },
}
\begin{tikzpicture}[
  >={Stealth[length=1.6mm]},
  box/.style={draw,rounded corners=2pt,minimum width=8mm,minimum height=8mm,font=\small},
  enc/.style={box,fill=blue!6},
  comb/.style={box,fill=orange!12,minimum height=9mm,minimum width=10mm},
  inp/.style={font=\scriptsize},
  lbl/.style={font=\scriptsize,text=black!70},
  every node/.style={inner sep=1.5pt}]
\node[inp] (a1) {$x_1$};
\node[inp,right=1.2mm of a1] (a2) {$\cdots$};
\node[inp,right=1.2mm of a2] (a3) {$x_{n}$};
\node[inp,right=5.5mm of a3] (b1) {$y_{1}$};
\node[inp,right=1.2mm of b1] (b2) {$\cdots$};
\node[inp,right=1.2mm of b2] (b3) {$y_{n}$};
\draw[decorate,decoration={brace,amplitude=4pt},black!55]
  ($(a1.north west)+(-1mm,1mm)$) -- ($(a3.north east)+(1mm,1mm)$)
  node[midway,above=3pt,lbl] {$A$};
\draw[decorate,decoration={brace,amplitude=4pt},black!55]
  ($(b1.north west)+(-1mm,1mm)$) -- ($(b3.north east)+(1mm,1mm)$)
  node[midway,above=3pt,lbl] {$B$};
\node[enc,below=6.5mm of a2] (pa) {$\vec\phi_A$};
\node[enc] (pb) at (pa -| b2) {$\vec\phi_B$};
\draw[->] (a1.south) -- ([xshift=-2.6mm]pa.north);
\draw[->] (a2.south) -- (pa.north);
\draw[->] (a3.south) -- ([xshift=2.6mm]pa.north);
\draw[->] (b1.south) -- ([xshift=-2.6mm]pb.north);
\draw[->] (b2.south) -- (pb.north);
\draw[->] (b3.south) -- ([xshift=2.6mm]pb.north);
\coordinate (mid) at ($(pa)!0.5!(pb)$);
\node[lbl] at ($(mid)+(0,-6.5mm)$) {encoders};
\node[comb] (g) at ($(mid)+(0,-15mm)$) {$g$};
\draw[->,thick] (pa.south) -- node[lbl,pos=0.45,left=2pt] {$\C^{K}$} (g.north west);
\draw[->,thick] (pb.south) -- (g.north east);
\node[inp,below=5mm of g,font=\small] (out) {$f(\vec x)$};
\draw[->,thick] (g.south) -- (out.north);
\node[lbl,anchor=west] at ($(g.south)+(1.8mm,-2.8mm)$) {decoder};
\begin{axis}[heat, clip=false,
  at={($(mid)+(3.00cm,-2.51cm)$)}, anchor=south west,
  xlabel={$K$}, ylabel={$p$},
  xmin=0, xmax=112, xtick={1,25,50,75,100},
  ymin=0, ymax=21.5, ytick={1,5,10,15,20},
  point meta min=0, point meta max=23,
]
\addplot[matrix plot*, mesh/cols=100, point meta=explicit] table[meta=m] {
x y m
1 1 0.3010
2 1 0.4771
3 1 0.6021
4 1 0.6990
5 1 0.7782
6 1 0.8451
7 1 0.9031
8 1 0.9542
9 1 1.0000
10 1 1.0414
11 1 1.0792
12 1 1.1139
13 1 1.1461
14 1 1.1761
15 1 1.2041
16 1 1.2304
17 1 1.2553
18 1 1.2788
19 1 1.3010
20 1 1.3222
21 1 1.3424
22 1 1.3617
23 1 1.3802
24 1 1.3979
25 1 1.4150
26 1 1.4314
27 1 1.4472
28 1 1.4624
29 1 1.4771
30 1 1.4914
31 1 1.5051
32 1 1.5185
33 1 1.5315
34 1 1.5441
35 1 1.5563
36 1 1.5682
37 1 1.5798
38 1 1.5911
39 1 1.6021
40 1 1.6128
41 1 1.6232
42 1 1.6335
43 1 1.6435
44 1 1.6532
45 1 1.6628
46 1 1.6721
47 1 1.6812
48 1 1.6902
49 1 1.6990
50 1 1.7076
51 1 1.7160
52 1 1.7243
53 1 1.7324
54 1 1.7404
55 1 1.7482
56 1 1.7559
57 1 1.7634
58 1 1.7709
59 1 1.7782
60 1 1.7853
61 1 1.7924
62 1 1.7993
63 1 1.8062
64 1 1.8129
65 1 1.8195
66 1 1.8261
67 1 1.8325
68 1 1.8388
69 1 1.8451
70 1 1.8513
71 1 1.8573
72 1 1.8633
73 1 1.8692
74 1 1.8751
75 1 1.8808
76 1 1.8865
77 1 1.8921
78 1 1.8976
79 1 1.9031
80 1 1.9085
81 1 1.9138
82 1 1.9191
83 1 1.9243
84 1 1.9294
85 1 1.9345
86 1 1.9395
87 1 1.9445
88 1 1.9494
89 1 1.9542
90 1 1.9590
91 1 1.9638
92 1 1.9685
93 1 1.9731
94 1 1.9777
95 1 1.9823
96 1 1.9868
97 1 1.9912
98 1 1.9956
99 1 2.0000
100 1 2.0043
1 2 0.4771
2 2 0.7782
3 2 1.0000
4 2 1.1761
5 2 1.3222
6 2 1.4472
7 2 1.5563
8 2 1.6532
9 2 1.7404
10 2 1.8195
11 2 1.8921
12 2 1.9590
13 2 2.0212
14 2 2.0792
15 2 2.1335
16 2 2.1847
17 2 2.2330
18 2 2.2788
19 2 2.3222
20 2 2.3636
21 2 2.4031
22 2 2.4409
23 2 2.4771
24 2 2.5119
25 2 2.5453
26 2 2.5775
27 2 2.6085
28 2 2.6385
29 2 2.6675
30 2 2.6955
31 2 2.7226
32 2 2.7490
33 2 2.7745
34 2 2.7993
35 2 2.8235
36 2 2.8470
37 2 2.8698
38 2 2.8921
39 2 2.9138
40 2 2.9350
41 2 2.9557
42 2 2.9759
43 2 2.9956
44 2 3.0149
45 2 3.0338
46 2 3.0523
47 2 3.0704
48 2 3.0881
49 2 3.1055
50 2 3.1225
51 2 3.1392
52 2 3.1556
53 2 3.1717
54 2 3.1875
55 2 3.2030
56 2 3.2183
57 2 3.2333
58 2 3.2480
59 2 3.2625
60 2 3.2767
61 2 3.2907
62 2 3.3045
63 2 3.3181
64 2 3.3314
65 2 3.3446
66 2 3.3576
67 2 3.3703
68 2 3.3829
69 2 3.3953
70 2 3.4076
71 2 3.4196
72 2 3.4315
73 2 3.4433
74 2 3.4548
75 2 3.4663
76 2 3.4776
77 2 3.4887
78 2 3.4997
79 2 3.5105
80 2 3.5213
81 2 3.5319
82 2 3.5423
83 2 3.5527
84 2 3.5629
85 2 3.5730
86 2 3.5830
87 2 3.5928
88 2 3.6026
89 2 3.6123
90 2 3.6218
91 2 3.6312
92 2 3.6406
93 2 3.6498
94 2 3.6590
95 2 3.6680
96 2 3.6770
97 2 3.6858
98 2 3.6946
99 2 3.7033
100 2 3.7119
1 3 0.6021
2 3 1.0000
3 3 1.3010
4 3 1.5441
5 3 1.7482
6 3 1.9243
7 3 2.0792
8 3 2.2175
9 3 2.3424
10 3 2.4564
11 3 2.5611
12 3 2.6580
13 3 2.7482
14 3 2.8325
15 3 2.9117
16 3 2.9863
17 3 3.0569
18 3 3.1239
19 3 3.1875
20 3 3.2482
21 3 3.3062
22 3 3.3617
23 3 3.4150
24 3 3.4661
25 3 3.5153
26 3 3.5628
27 3 3.6085
28 3 3.6527
29 3 3.6955
30 3 3.7369
31 3 3.7770
32 3 3.8159
33 3 3.8537
34 3 3.8904
35 3 3.9261
36 3 3.9609
37 3 3.9948
38 3 4.0278
39 3 4.0599
40 3 4.0914
41 3 4.1220
42 3 4.1520
43 3 4.1813
44 3 4.2099
45 3 4.2379
46 3 4.2654
47 3 4.2923
48 3 4.3186
49 3 4.3444
50 3 4.3697
51 3 4.3945
52 3 4.4189
53 3 4.4428
54 3 4.4663
55 3 4.4893
56 3 4.5120
57 3 4.5343
58 3 4.5562
59 3 4.5777
60 3 4.5989
61 3 4.6198
62 3 4.6403
63 3 4.6605
64 3 4.6804
65 3 4.7000
66 3 4.7193
67 3 4.7383
68 3 4.7571
69 3 4.7755
70 3 4.7938
71 3 4.8117
72 3 4.8295
73 3 4.8470
74 3 4.8642
75 3 4.8812
76 3 4.8981
77 3 4.9147
78 3 4.9311
79 3 4.9472
80 3 4.9632
81 3 4.9790
82 3 4.9946
83 3 5.0100
84 3 5.0253
85 3 5.0403
86 3 5.0552
87 3 5.0700
88 3 5.0845
89 3 5.0989
90 3 5.1132
91 3 5.1272
92 3 5.1412
93 3 5.1550
94 3 5.1686
95 3 5.1821
96 3 5.1955
97 3 5.2087
98 3 5.2218
99 3 5.2348
100 3 5.2476
1 4 0.6990
2 4 1.1761
3 4 1.5441
4 4 1.8451
5 4 2.1004
6 4 2.3222
7 4 2.5185
8 4 2.6946
9 4 2.8543
10 4 3.0004
11 4 3.1351
12 4 3.2601
13 4 3.3766
14 4 3.4857
15 4 3.5884
16 4 3.6853
17 4 3.7771
18 4 3.8642
19 4 3.9472
20 4 4.0264
21 4 4.1021
22 4 4.1746
23 4 4.2443
24 4 4.3112
25 4 4.3757
26 4 4.4378
27 4 4.4978
28 4 4.5558
29 4 4.6119
30 4 4.6663
31 4 4.7190
32 4 4.7702
33 4 4.8198
34 4 4.8681
35 4 4.9151
36 4 4.9609
37 4 5.0055
38 4 5.0489
39 4 5.0914
40 4 5.1327
41 4 5.1732
42 4 5.2127
43 4 5.2513
44 4 5.2891
45 4 5.3261
46 4 5.3623
47 4 5.3978
48 4 5.4325
49 4 5.4666
50 4 5.5000
51 4 5.5328
52 4 5.5650
53 4 5.5966
54 4 5.6276
55 4 5.6581
56 4 5.6881
57 4 5.7175
58 4 5.7465
59 4 5.7750
60 4 5.8030
61 4 5.8306
62 4 5.8578
63 4 5.8845
64 4 5.9108
65 4 5.9368
66 4 5.9623
67 4 5.9875
68 4 6.0123
69 4 6.0368
70 4 6.0609
71 4 6.0847
72 4 6.1082
73 4 6.1314
74 4 6.1542
75 4 6.1768
76 4 6.1991
77 4 6.2211
78 4 6.2428
79 4 6.2643
80 4 6.2854
81 4 6.3064
82 4 6.3271
83 4 6.3475
84 4 6.3677
85 4 6.3877
86 4 6.4074
87 4 6.4269
88 4 6.4463
89 4 6.4653
90 4 6.4842
91 4 6.5029
92 4 6.5214
93 4 6.5397
94 4 6.5578
95 4 6.5757
96 4 6.5934
97 4 6.6110
98 4 6.6283
99 4 6.6455
100 4 6.6626
1 5 0.7782
2 5 1.3222
3 5 1.7482
4 5 2.1004
5 5 2.4014
6 5 2.6646
7 5 2.8987
8 5 3.1096
9 5 3.3015
10 5 3.4776
11 5 3.6403
12 5 3.7916
13 5 3.9329
14 5 4.0655
15 5 4.1904
16 5 4.3085
17 5 4.4205
18 5 4.5270
19 5 4.6284
20 5 4.7253
21 5 4.8181
22 5 4.9070
23 5 4.9925
24 5 5.0747
25 5 5.1538
26 5 5.2302
27 5 5.3040
28 5 5.3754
29 5 5.4444
30 5 5.5114
31 5 5.5763
32 5 5.6394
33 5 5.7007
34 5 5.7602
35 5 5.8182
36 5 5.8747
37 5 5.9298
38 5 5.9834
39 5 6.0358
40 5 6.0870
41 5 6.1370
42 5 6.1858
43 5 6.2336
44 5 6.2803
45 5 6.3261
46 5 6.3709
47 5 6.4148
48 5 6.4578
49 5 6.5000
50 5 6.5414
51 5 6.5820
52 5 6.6219
53 5 6.6611
54 5 6.6995
55 5 6.7373
56 5 6.7745
57 5 6.8110
58 5 6.8469
59 5 6.8822
60 5 6.9170
61 5 6.9512
62 5 6.9849
63 5 7.0180
64 5 7.0507
65 5 7.0829
66 5 7.1146
67 5 7.1459
68 5 7.1767
69 5 7.2071
70 5 7.2370
71 5 7.2666
72 5 7.2957
73 5 7.3245
74 5 7.3529
75 5 7.3809
76 5 7.4086
77 5 7.4359
78 5 7.4629
79 5 7.4896
80 5 7.5159
81 5 7.5419
82 5 7.5676
83 5 7.5930
84 5 7.6181
85 5 7.6430
86 5 7.6675
87 5 7.6918
88 5 7.7158
89 5 7.7395
90 5 7.7630
91 5 7.7862
92 5 7.8092
93 5 7.8319
94 5 7.8544
95 5 7.8767
96 5 7.8988
97 5 7.9206
98 5 7.9422
99 5 7.9636
100 5 7.9848
1 6 0.8451
2 6 1.4472
3 6 1.9243
4 6 2.3222
5 6 2.6646
6 6 2.9657
7 6 3.2345
8 6 3.4776
9 6 3.6994
10 6 3.9035
11 6 4.0926
12 6 4.2687
13 6 4.4335
14 6 4.5884
15 6 4.7345
16 6 4.8728
17 6 5.0041
18 6 5.1290
19 6 5.2482
20 6 5.3622
21 6 5.4713
22 6 5.5760
23 6 5.6767
24 6 5.7736
25 6 5.8670
26 6 5.9572
27 6 6.0444
28 6 6.1287
29 6 6.2104
30 6 6.2895
31 6 6.3664
32 6 6.4410
33 6 6.5136
34 6 6.5841
35 6 6.6529
36 6 6.7198
37 6 6.7851
38 6 6.8487
39 6 6.9109
40 6 6.9716
41 6 7.0309
42 6 7.0889
43 6 7.1456
44 6 7.2011
45 6 7.2555
46 6 7.3087
47 6 7.3609
48 6 7.4121
49 6 7.4622
50 6 7.5115
51 6 7.5598
52 6 7.6072
53 6 7.6538
54 6 7.6995
55 6 7.7445
56 6 7.7887
57 6 7.8322
58 6 7.8749
59 6 7.9170
60 6 7.9584
61 6 7.9991
62 6 8.0392
63 6 8.0787
64 6 8.1177
65 6 8.1560
66 6 8.1938
67 6 8.2310
68 6 8.2678
69 6 8.3040
70 6 8.3397
71 6 8.3749
72 6 8.4097
73 6 8.4440
74 6 8.4778
75 6 8.5113
76 6 8.5443
77 6 8.5769
78 6 8.6090
79 6 8.6408
80 6 8.6722
81 6 8.7033
82 6 8.7339
83 6 8.7643
84 6 8.7942
85 6 8.8238
86 6 8.8531
87 6 8.8821
88 6 8.9107
89 6 8.9391
90 6 8.9671
91 6 8.9948
92 6 9.0223
93 6 9.0494
94 6 9.0763
95 6 9.1029
96 6 9.1292
97 6 9.1553
98 6 9.1811
99 6 9.2066
100 6 9.2320
1 7 0.9031
2 7 1.5563
3 7 2.0792
4 7 2.5185
5 7 2.8987
6 7 3.2345
7 7 3.5355
8 7 3.8085
9 7 4.0584
10 7 4.2889
11 7 4.5028
12 7 4.7023
13 7 4.8894
14 7 5.0655
15 7 5.2318
16 7 5.3894
17 7 5.5392
18 7 5.6819
19 7 5.8181
20 7 5.9484
21 7 6.0734
22 7 6.1933
23 7 6.3087
24 7 6.4199
25 7 6.5271
26 7 6.6306
27 7 6.7308
28 7 6.8277
29 7 6.9216
30 7 7.0126
31 7 7.1011
32 7 7.1870
33 7 7.2705
34 7 7.3518
35 7 7.4310
36 7 7.5082
37 7 7.5834
38 7 7.6569
39 7 7.7286
40 7 7.7986
41 7 7.8670
42 7 7.9340
43 7 7.9995
44 7 8.0636
45 7 8.1264
46 7 8.1879
47 7 8.2482
48 7 8.3073
49 7 8.3653
50 7 8.4222
51 7 8.4781
52 7 8.5329
53 7 8.5868
54 7 8.6398
55 7 8.6918
56 7 8.7429
57 7 8.7932
58 7 8.8427
59 7 8.8914
60 7 8.9393
61 7 8.9865
62 7 9.0330
63 7 9.0787
64 7 9.1238
65 7 9.1682
66 7 9.2120
67 7 9.2552
68 7 9.2977
69 7 9.3397
70 7 9.3811
71 7 9.4219
72 7 9.4622
73 7 9.5020
74 7 9.5412
75 7 9.5800
76 7 9.6182
77 7 9.6560
78 7 9.6934
79 7 9.7302
80 7 9.7667
81 7 9.8027
82 7 9.8382
83 7 9.8734
84 7 9.9082
85 7 9.9425
86 7 9.9765
87 7 10.0101
88 7 10.0434
89 7 10.0762
90 7 10.1088
91 7 10.1410
92 7 10.1728
93 7 10.2043
94 7 10.2355
95 7 10.2664
96 7 10.2970
97 7 10.3272
98 7 10.3572
99 7 10.3869
100 7 10.4162
1 8 0.9542
2 8 1.6532
3 8 2.2175
4 8 2.6946
5 8 3.1096
6 8 3.4776
7 8 3.8085
8 8 4.1096
9 8 4.3858
10 8 4.6411
11 8 4.8784
12 8 5.1003
13 8 5.3085
14 8 5.5048
15 8 5.6905
16 8 5.8666
17 8 6.0341
18 8 6.1938
19 8 6.3464
20 8 6.4925
21 8 6.6327
22 8 6.7674
23 8 6.8970
24 8 7.0219
25 8 7.1425
26 8 7.2590
27 8 7.3717
28 8 7.4809
29 8 7.5867
30 8 7.6893
31 8 7.7890
32 8 7.8860
33 8 7.9802
34 8 8.0720
35 8 8.1614
36 8 8.2485
37 8 8.3336
38 8 8.4165
39 8 8.4976
40 8 8.5767
41 8 8.6542
42 8 8.7299
43 8 8.8040
44 8 8.8765
45 8 8.9476
46 8 9.0172
47 8 9.0855
48 8 9.1524
49 8 9.2181
50 8 9.2826
51 8 9.3459
52 8 9.4080
53 8 9.4691
54 8 9.5291
55 8 9.5880
56 8 9.6460
57 8 9.7031
58 8 9.7592
59 8 9.8144
60 8 9.8688
61 8 9.9223
62 8 9.9750
63 8 10.0269
64 8 10.0781
65 8 10.1285
66 8 10.1782
67 8 10.2271
68 8 10.2754
69 8 10.3231
70 8 10.3701
71 8 10.4165
72 8 10.4622
73 8 10.5074
74 8 10.5520
75 8 10.5960
76 8 10.6394
77 8 10.6824
78 8 10.7248
79 8 10.7667
80 8 10.8081
81 8 10.8490
82 8 10.8894
83 8 10.9294
84 8 10.9689
85 8 11.0079
86 8 11.0466
87 8 11.0848
88 8 11.1225
89 8 11.1599
90 8 11.1969
91 8 11.2335
92 8 11.2697
93 8 11.3056
94 8 11.3410
95 8 11.3761
96 8 11.4109
97 8 11.4453
98 8 11.4794
99 8 11.5132
100 8 11.5466
1 9 1.0000
2 9 1.7404
3 9 2.3424
4 9 2.8543
5 9 3.3015
6 9 3.6994
7 9 4.0584
8 9 4.3858
9 9 4.6868
10 9 4.9656
11 9 5.2252
12 9 5.4682
13 9 5.6967
14 9 5.9123
15 9 6.1164
16 9 6.3103
17 9 6.4948
18 9 6.6709
19 9 6.8393
20 9 7.0007
21 9 7.1556
22 9 7.3045
23 9 7.4479
24 9 7.5862
25 9 7.7198
26 9 7.8489
27 9 7.9738
28 9 8.0948
29 9 8.2122
30 9 8.3262
31 9 8.4369
32 9 8.5445
33 9 8.6492
34 9 8.7512
35 9 8.8506
36 9 8.9475
37 9 9.0421
38 9 9.1344
39 9 9.2246
40 9 9.3127
41 9 9.3989
42 9 9.4832
43 9 9.5657
44 9 9.6466
45 9 9.7257
46 9 9.8033
47 9 9.8794
48 9 9.9541
49 9 10.0273
50 9 10.0992
51 9 10.1698
52 9 10.2391
53 9 10.3072
54 9 10.3742
55 9 10.4400
56 9 10.5047
57 9 10.5684
58 9 10.6310
59 9 10.6927
60 9 10.7534
61 9 10.8131
62 9 10.8720
63 9 10.9300
64 9 10.9871
65 9 11.0435
66 9 11.0990
67 9 11.1537
68 9 11.2077
69 9 11.2609
70 9 11.3135
71 9 11.3653
72 9 11.4165
73 9 11.4669
74 9 11.5168
75 9 11.5660
76 9 11.6146
77 9 11.6626
78 9 11.7100
79 9 11.7569
80 9 11.8032
81 9 11.8490
82 9 11.8942
83 9 11.9389
84 9 11.9831
85 9 12.0268
86 9 12.0700
87 9 12.1128
88 9 12.1551
89 9 12.1969
90 9 12.2383
91 9 12.2793
92 9 12.3198
93 9 12.3599
94 9 12.3996
95 9 12.4389
96 9 12.4779
97 9 12.5164
98 9 12.5545
99 9 12.5923
100 9 12.6298
1 10 1.0414
2 10 1.8195
3 10 2.4564
4 10 3.0004
5 10 3.4776
6 10 3.9035
7 10 4.2889
8 10 4.6411
9 10 4.9656
10 10 5.2666
11 10 5.5474
12 10 5.8107
13 10 6.0585
14 10 6.2925
15 10 6.5144
16 10 6.7252
17 10 6.9262
18 10 7.1180
19 10 7.3017
20 10 7.4778
21 10 7.6469
22 10 7.8096
23 10 7.9664
24 10 8.1177
25 10 8.2638
26 10 8.4052
27 10 8.5420
28 10 8.6746
29 10 8.8033
30 10 8.9282
31 10 9.0496
32 10 9.1677
33 10 9.2827
34 10 9.3947
35 10 9.5038
36 10 9.6103
37 10 9.7142
38 10 9.8156
39 10 9.9148
40 10 10.0117
41 10 10.1065
42 10 10.1992
43 10 10.2900
44 10 10.3790
45 10 10.4661
46 10 10.5515
47 10 10.6353
48 10 10.7175
49 10 10.7982
50 10 10.8773
51 10 10.9551
52 10 11.0315
53 10 11.1065
54 10 11.1803
55 10 11.2529
56 10 11.3242
57 10 11.3944
58 10 11.4635
59 10 11.5315
60 10 11.5985
61 10 11.6644
62 10 11.7293
63 10 11.7933
64 10 11.8564
65 10 11.9185
66 10 11.9798
67 10 12.0402
68 10 12.0998
69 10 12.1586
70 10 12.2166
71 10 12.2738
72 10 12.3303
73 10 12.3860
74 10 12.4411
75 10 12.4954
76 10 12.5491
77 10 12.6021
78 10 12.6545
79 10 12.7063
80 10 12.7574
81 10 12.8080
82 10 12.8580
83 10 12.9074
84 10 12.9562
85 10 13.0045
86 10 13.0523
87 10 13.0996
88 10 13.1463
89 10 13.1925
90 10 13.2383
91 10 13.2836
92 10 13.3284
93 10 13.3728
94 10 13.4167
95 10 13.4601
96 10 13.5032
97 10 13.5458
98 10 13.5880
99 10 13.6298
100 10 13.6712
1 11 1.0792
2 11 1.8921
3 11 2.5611
4 11 3.1351
5 11 3.6403
6 11 4.0926
7 11 4.5028
8 11 4.8784
9 11 5.2252
10 11 5.5474
11 11 5.8485
12 11 6.1310
13 11 6.3973
14 11 6.6491
15 11 6.8880
16 11 7.1152
17 11 7.3319
18 11 7.5390
19 11 7.7374
20 11 7.9277
21 11 8.1107
22 11 8.2868
23 11 8.4565
24 11 8.6204
25 11 8.7787
26 11 8.9320
27 11 9.0804
28 11 9.2243
29 11 9.3640
30 11 9.4996
31 11 9.6315
32 11 9.7598
33 11 9.8848
34 11 10.0065
35 11 10.1252
36 11 10.2410
37 11 10.3540
38 11 10.4644
39 11 10.5723
40 11 10.6778
41 11 10.7811
42 11 10.8821
43 11 10.9810
44 11 11.0779
45 11 11.1729
46 11 11.2660
47 11 11.3573
48 11 11.4470
49 11 11.5349
50 11 11.6213
51 11 11.7061
52 11 11.7894
53 11 11.8713
54 11 11.9519
55 11 12.0310
56 11 12.1089
57 11 12.1856
58 11 12.2610
59 11 12.3352
60 11 12.4083
61 11 12.4803
62 11 12.5513
63 11 12.6212
64 11 12.6900
65 11 12.7579
66 11 12.8249
67 11 12.8909
68 11 12.9560
69 11 13.0203
70 11 13.0837
71 11 13.1462
72 11 13.2080
73 11 13.2689
74 11 13.3291
75 11 13.3885
76 11 13.4472
77 11 13.5052
78 11 13.5625
79 11 13.6191
80 11 13.6751
81 11 13.7304
82 11 13.7851
83 11 13.8391
84 11 13.8926
85 11 13.9454
86 11 13.9977
87 11 14.0494
88 11 14.1005
89 11 14.1512
90 11 14.2012
91 11 14.2508
92 11 14.2998
93 11 14.3484
94 11 14.3965
95 11 14.4440
96 11 14.4911
97 11 14.5378
98 11 14.5840
99 11 14.6298
100 11 14.6751
1 12 1.1139
2 12 1.9590
3 12 2.6580
4 12 3.2601
5 12 3.7916
6 12 4.2687
7 12 4.7023
8 12 5.1003
9 12 5.4682
10 12 5.8107
11 12 6.1310
12 12 6.4320
13 12 6.7160
14 12 6.9849
15 12 7.2401
16 12 7.4832
17 12 7.7151
18 12 7.9370
19 12 8.1496
20 12 8.3537
21 12 8.5500
22 12 8.7391
23 12 8.9214
24 12 9.0975
25 12 9.2678
26 12 9.4326
27 12 9.5923
28 12 9.7472
29 12 9.8976
30 12 10.0437
31 12 10.1858
32 12 10.3241
33 12 10.4588
34 12 10.5901
35 12 10.7181
36 12 10.8430
37 12 10.9650
38 12 11.0842
39 12 11.2007
40 12 11.3147
41 12 11.4262
42 12 11.5353
43 12 11.6422
44 12 11.7469
45 12 11.8496
46 12 11.9503
47 12 12.0490
48 12 12.1459
49 12 12.2411
50 12 12.3345
51 12 12.4263
52 12 12.5164
53 12 12.6051
54 12 12.6922
55 12 12.7779
56 12 12.8623
57 12 12.9452
58 12 13.0269
59 12 13.1073
60 12 13.1865
61 12 13.2645
62 12 13.3413
63 12 13.4170
64 12 13.4917
65 12 13.5652
66 12 13.6378
67 12 13.7094
68 12 13.7799
69 12 13.8496
70 12 13.9183
71 12 13.9861
72 12 14.0531
73 12 14.1191
74 12 14.1844
75 12 14.2489
76 12 14.3125
77 12 14.3754
78 12 14.4376
79 12 14.4990
80 12 14.5597
81 12 14.6197
82 12 14.6790
83 12 14.7377
84 12 14.7956
85 12 14.8530
86 12 14.9097
87 12 14.9658
88 12 15.0214
89 12 15.0763
90 12 15.1307
91 12 15.1844
92 12 15.2377
93 12 15.2904
94 12 15.3426
95 12 15.3942
96 12 15.4454
97 12 15.4960
98 12 15.5462
99 12 15.5959
100 12 15.6451
1 13 1.1461
2 13 2.0212
3 13 2.7482
4 13 3.3766
5 13 3.9329
6 13 4.4335
7 13 4.8894
8 13 5.3085
9 13 5.6967
10 13 6.0585
11 13 6.3973
12 13 6.7160
13 13 7.0171
14 13 7.3023
15 13 7.5734
16 13 7.8316
17 13 8.0783
18 13 8.3144
19 13 8.5408
20 13 8.7583
21 13 8.9675
22 13 9.1692
23 13 9.3638
24 13 9.5518
25 13 9.7336
26 13 9.9097
27 13 10.0804
28 13 10.2460
29 13 10.4069
30 13 10.5632
31 13 10.7153
32 13 10.8634
33 13 11.0076
34 13 11.1482
35 13 11.2854
36 13 11.4193
37 13 11.5501
38 13 11.6778
39 13 11.8028
40 13 11.9250
41 13 12.0446
42 13 12.1617
43 13 12.2764
44 13 12.3889
45 13 12.4991
46 13 12.6072
47 13 12.7132
48 13 12.8173
49 13 12.9195
50 13 13.0199
51 13 13.1185
52 13 13.2154
53 13 13.3107
54 13 13.4044
55 13 13.4965
56 13 13.5872
57 13 13.6764
58 13 13.7642
59 13 13.8507
60 13 13.9359
61 13 14.0198
62 13 14.1024
63 13 14.1839
64 13 14.2642
65 13 14.3434
66 13 14.4215
67 13 14.4985
68 13 14.5745
69 13 14.6494
70 13 14.7234
71 13 14.7964
72 13 14.8685
73 13 14.9397
74 13 15.0100
75 13 15.0794
76 13 15.1480
77 13 15.2157
78 13 15.2827
79 13 15.3488
80 13 15.4142
81 13 15.4789
82 13 15.5428
83 13 15.6060
84 13 15.6685
85 13 15.7303
86 13 15.7914
87 13 15.8519
88 13 15.9117
89 13 15.9710
90 13 16.0295
91 13 16.0875
92 13 16.1449
93 13 16.2018
94 13 16.2580
95 13 16.3137
96 13 16.3689
97 13 16.4235
98 13 16.4776
99 13 16.5312
100 13 16.5843
1 14 1.1761
2 14 2.0792
3 14 2.8325
4 14 3.4857
5 14 4.0655
6 14 4.5884
7 14 5.0655
8 14 5.5048
9 14 5.9123
10 14 6.2925
11 14 6.6491
12 14 6.9849
13 14 7.3023
14 14 7.6033
15 14 7.8896
16 14 8.1626
17 14 8.4235
18 14 8.6734
19 14 8.9132
20 14 9.1436
21 14 9.3655
22 14 9.5794
23 14 9.7858
24 14 9.9854
25 14 10.1785
26 14 10.3656
27 14 10.5470
28 14 10.7231
29 14 10.8942
30 14 11.0605
31 14 11.2224
32 14 11.3800
33 14 11.5336
34 14 11.6833
35 14 11.8295
36 14 11.9721
37 14 12.1115
38 14 12.2477
39 14 12.3809
40 14 12.5113
41 14 12.6388
42 14 12.7638
43 14 12.8862
44 14 13.0062
45 14 13.1238
46 14 13.2392
47 14 13.3524
48 14 13.4636
49 14 13.5727
50 14 13.6799
51 14 13.7853
52 14 13.8888
53 14 13.9906
54 14 14.0907
55 14 14.1892
56 14 14.2861
57 14 14.3815
58 14 14.4754
59 14 14.5679
60 14 14.6590
61 14 14.7487
62 14 14.8371
63 14 14.9243
64 14 15.0102
65 14 15.0949
66 14 15.1784
67 14 15.2609
68 14 15.3422
69 14 15.4224
70 14 15.5016
71 14 15.5797
72 14 15.6569
73 14 15.7331
74 14 15.8083
75 14 15.8827
76 14 15.9561
77 14 16.0287
78 14 16.1003
79 14 16.1712
80 14 16.2412
81 14 16.3105
82 14 16.3789
83 14 16.4466
84 14 16.5136
85 14 16.5798
86 14 16.6453
87 14 16.7101
88 14 16.7742
89 14 16.8377
90 14 16.9005
91 14 16.9626
92 14 17.0241
93 14 17.0850
94 14 17.1453
95 14 17.2050
96 14 17.2641
97 14 17.3227
98 14 17.3807
99 14 17.4381
100 14 17.4950
1 15 1.2041
2 15 2.1335
3 15 2.9117
4 15 3.5884
5 15 4.1904
6 15 4.7345
7 15 5.2318
8 15 5.6905
9 15 6.1164
10 15 6.5144
11 15 6.8880
12 15 7.2401
13 15 7.5734
14 15 7.8896
15 15 8.1907
16 15 8.4779
17 15 8.7526
18 15 9.0158
19 15 9.2686
20 15 9.5116
21 15 9.7457
22 15 9.9715
23 15 10.1895
24 15 10.4004
25 15 10.6045
26 15 10.8023
27 15 10.9942
28 15 11.1805
29 15 11.3616
30 15 11.5377
31 15 11.7090
32 15 11.8760
33 15 12.0387
34 15 12.1974
35 15 12.3523
36 15 12.5036
37 15 12.6514
38 15 12.7959
39 15 12.9372
40 15 13.0755
41 15 13.2109
42 15 13.3436
43 15 13.4735
44 15 13.6009
45 15 13.7259
46 15 13.8484
47 15 13.9687
48 15 14.0868
49 15 14.2028
50 15 14.3168
51 15 14.4287
52 15 14.5388
53 15 14.6470
54 15 14.7535
55 15 14.8582
56 15 14.9613
57 15 15.0628
58 15 15.1626
59 15 15.2610
60 15 15.3579
61 15 15.4534
62 15 15.5475
63 15 15.6403
64 15 15.7317
65 15 15.8219
66 15 15.9108
67 15 15.9986
68 15 16.0851
69 15 16.1706
70 15 16.2549
71 15 16.3381
72 15 16.4203
73 15 16.5015
74 15 16.5816
75 15 16.6608
76 15 16.7391
77 15 16.8163
78 15 16.8927
79 15 16.9682
80 15 17.0429
81 15 17.1167
82 15 17.1896
83 15 17.2618
84 15 17.3331
85 15 17.4037
86 15 17.4735
87 15 17.5426
88 15 17.6110
89 15 17.6786
90 15 17.7455
91 15 17.8118
92 15 17.8774
93 15 17.9424
94 15 18.0066
95 15 18.0703
96 15 18.1334
97 15 18.1958
98 15 18.2577
99 15 18.3189
100 15 18.3796
1 16 1.2304
2 16 2.1847
3 16 2.9863
4 16 3.6853
5 16 4.3085
6 16 4.8728
7 16 5.3894
8 16 5.8666
9 16 6.3103
10 16 6.7252
11 16 7.1152
12 16 7.4832
13 16 7.8316
14 16 8.1626
15 16 8.4779
16 16 8.7789
17 16 9.0670
18 16 9.3432
19 16 9.6085
20 16 9.8638
21 16 10.1098
22 16 10.3471
23 16 10.5765
24 16 10.7983
25 16 11.0132
26 16 11.2214
27 16 11.4235
28 16 11.6198
29 16 11.8107
30 16 11.9963
31 16 12.1770
32 16 12.3531
33 16 12.5248
34 16 12.6923
35 16 12.8558
36 16 13.0155
37 16 13.1716
38 16 13.3242
39 16 13.4735
40 16 13.6196
41 16 13.7627
42 16 13.9029
43 16 14.0403
44 16 14.1750
45 16 14.3071
46 16 14.4367
47 16 14.5639
48 16 14.6889
49 16 14.8116
50 16 14.9322
51 16 15.0507
52 16 15.1672
53 16 15.2818
54 16 15.3945
55 16 15.5054
56 16 15.6145
57 16 15.7220
58 16 15.8278
59 16 15.9320
60 16 16.0346
61 16 16.1358
62 16 16.2355
63 16 16.3338
64 16 16.4307
65 16 16.5263
66 16 16.6205
67 16 16.7135
68 16 16.8053
69 16 16.8959
70 16 16.9853
71 16 17.0735
72 16 17.1607
73 16 17.2468
74 16 17.3318
75 16 17.4157
76 16 17.4987
77 16 17.5807
78 16 17.6617
79 16 17.7418
80 16 17.8210
81 16 17.8993
82 16 17.9767
83 16 18.0533
84 16 18.1290
85 16 18.2039
86 16 18.2780
87 16 18.3513
88 16 18.4239
89 16 18.4957
90 16 18.5667
91 16 18.6371
92 16 18.7067
93 16 18.7757
94 16 18.8439
95 16 18.9115
96 16 18.9785
97 16 19.0448
98 16 19.1105
99 16 19.1755
100 16 19.2400
1 17 1.2553
2 17 2.2330
3 17 3.0569
4 17 3.7771
5 17 4.4205
6 17 5.0041
7 17 5.5392
8 17 6.0341
9 17 6.4948
10 17 6.9262
11 17 7.3319
12 17 7.7151
13 17 8.0783
14 17 8.4235
15 17 8.7526
16 17 9.0670
17 17 9.3680
18 17 9.6568
19 17 9.9344
20 17 10.2015
21 17 10.4591
22 17 10.7077
23 17 10.9481
24 17 11.1807
25 17 11.4060
26 17 11.6245
27 17 11.8365
28 17 12.0426
29 17 12.2430
30 17 12.4379
31 17 12.6278
32 17 12.8129
33 17 12.9933
34 17 13.1694
35 17 13.3413
36 17 13.5093
37 17 13.6735
38 17 13.8341
39 17 13.9912
40 17 14.1450
41 17 14.2957
42 17 14.4433
43 17 14.5880
44 17 14.7298
45 17 14.8690
46 17 15.0056
47 17 15.1397
48 17 15.2714
49 17 15.4007
50 17 15.5278
51 17 15.6527
52 17 15.7756
53 17 15.8964
54 17 16.0153
55 17 16.1322
56 17 16.2474
57 17 16.3607
58 17 16.4724
59 17 16.5823
60 17 16.6907
61 17 16.7974
62 17 16.9027
63 17 17.0064
64 17 17.1087
65 17 17.2096
66 17 17.3092
67 17 17.4074
68 17 17.5043
69 17 17.5999
70 17 17.6943
71 17 17.7876
72 17 17.8796
73 17 17.9705
74 17 18.0604
75 17 18.1491
76 17 18.2368
77 17 18.3234
78 17 18.4090
79 17 18.4937
80 17 18.5773
81 17 18.6601
82 17 18.7419
83 17 18.8228
84 17 18.9029
85 17 18.9821
86 17 19.0604
87 17 19.1379
88 17 19.2146
89 17 19.2905
90 17 19.3657
91 17 19.4401
92 17 19.5137
93 17 19.5866
94 17 19.6588
95 17 19.7303
96 17 19.8011
97 17 19.8712
98 17 19.9407
99 17 20.0095
100 17 20.0777
1 18 1.2788
2 18 2.2788
3 18 3.1239
4 18 3.8642
5 18 4.5270
6 18 5.1290
7 18 5.6819
8 18 6.1938
9 18 6.6709
10 18 7.1180
11 18 7.5390
12 18 7.9370
13 18 8.3144
14 18 8.6734
15 18 9.0158
16 18 9.3432
17 18 9.6568
18 18 9.9579
19 18 10.2473
20 18 10.5261
21 18 10.7949
22 18 11.0545
23 18 11.3056
24 18 11.5486
25 18 11.7842
26 18 12.0126
27 18 12.2345
28 18 12.4501
29 18 12.6598
30 18 12.8639
31 18 13.0627
32 18 13.2566
33 18 13.4456
34 18 13.6301
35 18 13.8104
36 18 13.9864
37 18 14.1586
38 18 14.3270
39 18 14.4918
40 18 14.6532
41 18 14.8113
42 18 14.9662
43 18 15.1180
44 18 15.2670
45 18 15.4131
46 18 15.5565
47 18 15.6973
48 18 15.8356
49 18 15.9715
50 18 16.1050
51 18 16.2363
52 18 16.3654
53 18 16.4924
54 18 16.6173
55 18 16.7403
56 18 16.8613
57 18 16.9805
58 18 17.0979
59 18 17.2136
60 18 17.3275
61 18 17.4398
62 18 17.5505
63 18 17.6596
64 18 17.7673
65 18 17.8734
66 18 17.9782
67 18 18.0815
68 18 18.1835
69 18 18.2842
70 18 18.3836
71 18 18.4817
72 18 18.5786
73 18 18.6743
74 18 18.7689
75 18 18.8623
76 18 18.9546
77 18 19.0458
78 18 19.1360
79 18 19.2252
80 18 19.3133
81 18 19.4004
82 18 19.4866
83 18 19.5719
84 18 19.6562
85 18 19.7396
86 18 19.8222
87 18 19.9038
88 18 19.9846
89 18 20.0646
90 18 20.1438
91 18 20.2222
92 18 20.2998
93 18 20.3767
94 18 20.4527
95 18 20.5281
96 18 20.6027
97 18 20.6767
98 18 20.7499
99 18 20.8224
100 18 20.8943
1 19 1.3010
2 19 2.3222
3 19 3.1875
4 19 3.9472
5 19 4.6284
6 19 5.2482
7 19 5.8181
8 19 6.3464
9 19 6.8393
10 19 7.3017
11 19 7.7374
12 19 8.1496
13 19 8.5408
14 19 8.9132
15 19 9.2686
16 19 9.6085
17 19 9.9344
18 19 10.2473
19 19 10.5483
20 19 10.8384
21 19 11.1182
22 19 11.3886
23 19 11.6501
24 19 11.9033
25 19 12.1489
26 19 12.3871
27 19 12.6185
28 19 12.8434
29 19 13.0623
30 19 13.2754
31 19 13.4830
32 19 13.6854
33 19 13.8829
34 19 14.0757
35 19 14.2640
36 19 14.4481
37 19 14.6280
38 19 14.8041
39 19 14.9765
40 19 15.1453
41 19 15.3107
42 19 15.4727
43 19 15.6317
44 19 15.7875
45 19 15.9405
46 19 16.0907
47 19 16.2381
48 19 16.3829
49 19 16.5253
50 19 16.6651
51 19 16.8027
52 19 16.9379
53 19 17.0710
54 19 17.2019
55 19 17.3308
56 19 17.4576
57 19 17.5826
58 19 17.7056
59 19 17.8269
60 19 17.9464
61 19 18.0641
62 19 18.1802
63 19 18.2947
64 19 18.4076
65 19 18.5190
66 19 18.6288
67 19 18.7373
68 19 18.8443
69 19 18.9499
70 19 19.0542
71 19 19.1572
72 19 19.2589
73 19 19.3594
74 19 19.4586
75 19 19.5567
76 19 19.6536
77 19 19.7494
78 19 19.8440
79 19 19.9376
80 19 20.0302
81 19 20.1217
82 19 20.2122
83 19 20.3017
84 19 20.3903
85 19 20.4779
86 19 20.5646
87 19 20.6504
88 19 20.7353
89 19 20.8193
90 19 20.9025
91 19 20.9848
92 19 21.0664
93 19 21.1471
94 19 21.2271
95 19 21.3062
96 19 21.3847
97 19 21.4624
98 19 21.5393
99 19 21.6156
100 19 21.6911
1 20 1.3222
2 20 2.3636
3 20 3.2482
4 20 4.0264
5 20 4.7253
6 20 5.3622
7 20 5.9484
8 20 6.4925
9 20 7.0007
10 20 7.4778
11 20 7.9277
12 20 8.3537
13 20 8.7583
14 20 9.1436
15 20 9.5116
16 20 9.8638
17 20 10.2015
18 20 10.5261
19 20 10.8384
20 20 11.1394
21 20 11.4300
22 20 11.7108
23 20 11.9825
24 20 12.2458
25 20 12.5010
26 20 12.7488
27 20 12.9896
28 20 13.2236
29 20 13.4514
30 20 13.6733
31 20 13.8895
32 20 14.1004
33 20 14.3061
34 20 14.5070
35 20 14.7033
36 20 14.8952
37 20 15.0829
38 20 15.2665
39 20 15.4463
40 20 15.6224
41 20 15.7950
42 20 15.9641
43 20 16.1300
44 20 16.2927
45 20 16.4524
46 20 16.6092
47 20 16.7632
48 20 16.9144
49 20 17.0631
50 20 17.2092
51 20 17.3529
52 20 17.4942
53 20 17.6333
54 20 17.7701
55 20 17.9048
56 20 18.0374
57 20 18.1680
58 20 18.2967
59 20 18.4235
60 20 18.5484
61 20 18.6716
62 20 18.7930
63 20 18.9127
64 20 19.0308
65 20 19.1473
66 20 19.2623
67 20 19.3757
68 20 19.4877
69 20 19.5983
70 20 19.7074
71 20 19.8152
72 20 19.9216
73 20 20.0268
74 20 20.1307
75 20 20.2334
76 20 20.3348
77 20 20.4351
78 20 20.5342
79 20 20.6322
80 20 20.7292
81 20 20.8250
82 20 20.9198
83 20 21.0135
84 20 21.1063
85 20 21.1981
86 20 21.2889
87 20 21.3787
88 20 21.4677
89 20 21.5557
90 20 21.6429
91 20 21.7291
92 20 21.8146
93 20 21.8992
94 20 21.9829
95 20 22.0659
96 20 22.1481
97 20 22.2295
98 20 22.3102
99 20 22.3901
100 20 22.4693
};
\addplot[black!55, line width=0.55pt, line join=round] coordinates {(100.00,2.83) (92.21,2.90) (82.35,3.00) (74.11,3.10) (67.17,3.20) (61.27,3.30) (56.21,3.40) (51.84,3.50) (48.04,3.60) (44.71,3.70) (41.78,3.80) (39.18,3.90) (36.88,4.00) (34.81,4.10) (32.96,4.20) (31.29,4.30) (29.77,4.40) (28.40,4.50) (27.15,4.60) (26.00,4.70) (24.95,4.80) (23.98,4.90) (23.09,5.00) (22.27,5.10) (21.50,5.20) (20.79,5.30) (20.13,5.40) (19.51,5.50) (18.94,5.60) (18.40,5.70) (17.89,5.80) (17.42,5.90) (16.97,6.00) (16.55,6.10) (16.15,6.20) (15.77,6.30) (15.41,6.40) (15.08,6.50) (14.76,6.60) (14.45,6.70) (14.16,6.80) (13.88,6.90) (13.62,7.00) (13.37,7.10) (13.13,7.20) (12.90,7.30) (12.68,7.40) (12.47,7.50) (12.27,7.60) (12.08,7.70) (11.89,7.80) (11.71,7.90) (11.54,8.00) (11.37,8.10) (11.22,8.20) (11.06,8.30) (10.91,8.40) (10.77,8.50) (10.63,8.60) (10.50,8.70) (10.37,8.80) (10.25,8.90) (10.13,9.00) (10.01,9.10) (9.90,9.20) (9.79,9.30) (9.68,9.40) (9.58,9.50) (9.48,9.60) (9.39,9.70) (9.29,9.80) (9.20,9.90) (9.11,10.00) (9.02,10.10) (8.94,10.20) (8.86,10.30) (8.78,10.40) (8.70,10.50) (8.63,10.60) (8.55,10.70) (8.48,10.80) (8.41,10.90) (8.34,11.00) (8.28,11.10) (8.21,11.20) (8.15,11.30) (8.08,11.40) (8.02,11.50) (7.96,11.60) (7.91,11.70) (7.85,11.80) (7.79,11.90) (7.74,12.00) (7.69,12.10) (7.64,12.20) (7.58,12.30) (7.53,12.40) (7.49,12.50) (7.44,12.60) (7.39,12.70) (7.35,12.80) (7.30,12.90) (7.26,13.00) (7.21,13.10) (7.17,13.20) (7.13,13.30) (7.09,13.40) (7.05,13.50) (7.01,13.60) (6.97,13.70) (6.93,13.80) (6.89,13.90) (6.86,14.00) (6.82,14.10) (6.79,14.20) (6.75,14.30) (6.72,14.40) (6.68,14.50) (6.65,14.60) (6.62,14.70) (6.59,14.80) (6.55,14.90) (6.52,15.00) (6.49,15.10) (6.46,15.20) (6.43,15.30) (6.40,15.40) (6.38,15.50) (6.35,15.60) (6.32,15.70) (6.29,15.80) (6.27,15.90) (6.24,16.00) (6.21,16.10) (6.19,16.20) (6.16,16.30) (6.14,16.40) (6.11,16.50) (6.09,16.60) (6.06,16.70) (6.04,16.80) (6.02,16.90) (5.99,17.00) (5.97,17.10) (5.95,17.20) (5.93,17.30) (5.90,17.40) (5.88,17.50) (5.86,17.60) (5.84,17.70) (5.82,17.80) (5.80,17.90) (5.78,18.00) (5.76,18.10) (5.74,18.20) (5.72,18.30) (5.70,18.40) (5.68,18.50) (5.66,18.60) (5.64,18.70) (5.62,18.80) (5.61,18.90) (5.59,19.00) (5.57,19.10) (5.55,19.20) (5.54,19.30) (5.52,19.40) (5.50,19.50) (5.49,19.60) (5.47,19.70) (5.45,19.80) (5.44,19.90) (5.42,20.00)};
\addplot[black!55, line width=0.55pt, line join=round] coordinates {(100.00,6.64) (97.64,6.70) (93.73,6.80) (90.09,6.90) (86.70,7.00) (83.53,7.10) (80.57,7.20) (77.79,7.30) (75.19,7.40) (72.74,7.50) (70.44,7.60) (68.28,7.70) (66.23,7.80) (64.30,7.90) (62.48,8.00) (60.75,8.10) (59.12,8.20) (57.56,8.30) (56.09,8.40) (54.69,8.50) (53.36,8.60) (52.09,8.70) (50.88,8.80) (49.73,8.90) (48.62,9.00) (47.57,9.10) (46.57,9.20) (45.60,9.30) (44.68,9.40) (43.80,9.50) (42.95,9.60) (42.13,9.70) (41.35,9.80) (40.60,9.90) (39.88,10.00) (39.18,10.10) (38.51,10.20) (37.87,10.30) (37.25,10.40) (36.65,10.50) (36.07,10.60) (35.51,10.70) (34.97,10.80) (34.45,10.90) (33.95,11.00) (33.46,11.10) (32.99,11.20) (32.53,11.30) (32.09,11.40) (31.66,11.50) (31.24,11.60) (30.84,11.70) (30.45,11.80) (30.07,11.90) (29.70,12.00) (29.34,12.10) (28.99,12.20) (28.65,12.30) (28.32,12.40) (28.00,12.50) (27.69,12.60) (27.39,12.70) (27.09,12.80) (26.81,12.90) (26.53,13.00) (26.25,13.10) (25.99,13.20) (25.73,13.30) (25.47,13.40) (25.23,13.50) (24.98,13.60) (24.75,13.70) (24.52,13.80) (24.29,13.90) (24.07,14.00) (23.86,14.10) (23.65,14.20) (23.45,14.30) (23.24,14.40) (23.05,14.50) (22.86,14.60) (22.67,14.70) (22.49,14.80) (22.31,14.90) (22.13,15.00) (21.96,15.10) (21.79,15.20) (21.62,15.30) (21.46,15.40) (21.30,15.50) (21.14,15.60) (20.99,15.70) (20.84,15.80) (20.69,15.90) (20.55,16.00) (20.41,16.10) (20.27,16.20) (20.13,16.30) (20.00,16.40) (19.87,16.50) (19.74,16.60) (19.61,16.70) (19.49,16.80) (19.36,16.90) (19.24,17.00) (19.12,17.10) (19.01,17.20) (18.89,17.30) (18.78,17.40) (18.67,17.50) (18.56,17.60) (18.45,17.70) (18.35,17.80) (18.25,17.90) (18.14,18.00) (18.04,18.10) (17.94,18.20) (17.85,18.30) (17.75,18.40) (17.66,18.50) (17.56,18.60) (17.47,18.70) (17.38,18.80) (17.29,18.90) (17.21,19.00) (17.12,19.10) (17.04,19.20) (16.95,19.30) (16.87,19.40) (16.79,19.50) (16.71,19.60) (16.63,19.70) (16.55,19.80) (16.47,19.90) (16.40,20.00)};
\addplot[black!55, line width=0.55pt, line join=round] coordinates {(100.00,11.33) (98.57,11.40) (96.57,11.50) (94.64,11.60) (92.78,11.70) (91.00,11.80) (89.27,11.90) (87.61,12.00) (86.01,12.10) (84.47,12.20) (82.98,12.30) (81.54,12.40) (80.15,12.50) (78.81,12.60) (77.51,12.70) (76.25,12.80) (75.03,12.90) (73.86,13.00) (72.72,13.10) (71.61,13.20) (70.54,13.30) (69.50,13.40) (68.49,13.50) (67.52,13.60) (66.57,13.70) (65.65,13.80) (64.75,13.90) (63.88,14.00) (63.04,14.10) (62.21,14.20) (61.41,14.30) (60.64,14.40) (59.88,14.50) (59.14,14.60) (58.42,14.70) (57.72,14.80) (57.04,14.90) (56.38,15.00) (55.73,15.10) (55.10,15.20) (54.49,15.30) (53.89,15.40) (53.30,15.50) (52.73,15.60) (52.17,15.70) (51.62,15.80) (51.09,15.90) (50.57,16.00) (50.06,16.10) (49.56,16.20) (49.08,16.30) (48.60,16.40) (48.14,16.50) (47.68,16.60) (47.24,16.70) (46.80,16.80) (46.38,16.90) (45.96,17.00) (45.55,17.10) (45.15,17.20) (44.76,17.30) (44.37,17.40) (44.00,17.50) (43.63,17.60) (43.27,17.70) (42.91,17.80) (42.56,17.90) (42.22,18.00) (41.89,18.10) (41.56,18.20) (41.24,18.30) (40.92,18.40) (40.61,18.50) (40.30,18.60) (40.00,18.70) (39.71,18.80) (39.42,18.90) (39.14,19.00) (38.86,19.10) (38.59,19.20) (38.32,19.30) (38.05,19.40) (37.79,19.50) (37.54,19.60) (37.29,19.70) (37.04,19.80) (36.79,19.90) (36.56,20.00)};
\addplot[black!55, line width=0.55pt, line join=round] coordinates {(100.00,16.91) (98.86,17.00) (97.68,17.10) (96.52,17.20) (95.39,17.30) (94.29,17.40) (93.21,17.50) (92.16,17.60) (91.13,17.70) (90.13,17.80) (89.15,17.90) (88.19,18.00) (87.25,18.10) (86.34,18.20) (85.44,18.30) (84.56,18.40) (83.70,18.50) (82.86,18.60) (82.04,18.70) (81.23,18.80) (80.45,18.90) (79.67,19.00) (78.92,19.10) (78.17,19.20) (77.45,19.30) (76.74,19.40) (76.04,19.50) (75.35,19.60) (74.68,19.70) (74.02,19.80) (73.38,19.90) (72.74,20.00)};
\addplot[black!55, dash pattern=on 1pt off 1.8pt, line width=0.9pt]
  coordinates {(1,1) (100,1)};
\node[font=\scriptsize, color=black!60, anchor=south west] at (axis cs:3,1.3) {Schmidt};
\node[font=\scriptsize, color=black!60, anchor=west] at (axis cs:101.5,2.85) {$10^{5}$};
\node[font=\scriptsize, color=black!60, anchor=west] at (axis cs:101.5,6.7) {$10^{10}$};
\node[font=\scriptsize, color=black!60, anchor=west] at (axis cs:101.5,11.4) {$10^{15}$};
\node[font=\scriptsize, color=black!60, anchor=west] at (axis cs:101.5,17.0) {$10^{20}$};
\end{axis}
\end{tikzpicture}
\caption{(Left) The general template for the neural encoder-decoder
architecture: the bipartitioned input $A|B$ is encoded by two encoders
$\vec\phi_{A,B}$ into a $K$-dimensional latent space and decoded, or
recombined, by $g(\vec\phi_A,\vec\phi_B)$. When $g$ is simply a bilinear
function, this coincides with the Schmidt decomposition for quantum
wavefunctions. When $g$ is nonlinear, the neural encoder-decoder architecture
can be far more expressive. We quantify this by the \textit{entangling
power}. (Right) Entangling power $\mathcal{E}_p(K)=\binom{K+p}{p}$
[Eq.~\eqref{eq:monomial_count}] as a function of latent width $K$ and degree
$p$ for polynomial decoders (colour on a logarithmic scale). Along the dotted
line $p=1$ (labelled Schmidt for bilinear decoders) the entangling power
grows only linearly, $\mathcal{E}_1(K)=K+1$, while modest degrees $p$ reach
astronomical count already at small width $K$.}
\label{fig:schematic}
\end{figure}
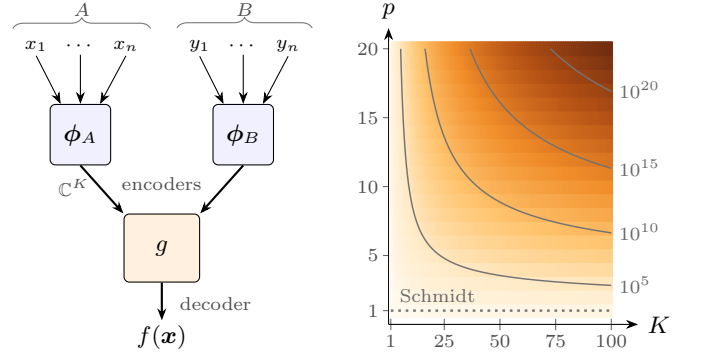

Broadly speaking, a neural network approximates a function by mapping input variables into hidden features, repeatedly mixing these features through linear and nonlinear transformations, and finally producing an output such as a wavefunction amplitude. Viewed across a cut $A|B$, this process can be idealized as two separate encoders that compress $x_A$ and $x_B$ into latent  vectors $\vec \phi_A$ and $\vec \phi_B$. These latent vectors are then recombined by a decoder $g:\mathbb C^K\times \mathbb C^K\to \mathbb C$
to produce the output (Fig.~\ref{fig:schematic}). The Schmidt decomposition is recovered when the decoder is restricted to a bilinear 
form. A generic neural decoder, however, is nonlinear in the latent variables. Thus the complexity of a neural representation across a cut is not fully captured by Schmidt rank alone. 

This observation motivates the central question of this work: what is the entangling power of a nonlinear encoder-decoder representation? By entangling power we mean the ability to generate large Schmidt rank across a bipartition from a small latent space by using a nonlinear decoder.

In this work, we quantify the entangling power of neural networks: $\mathcal{E}=e^{S_{\mathrm{max}}}$, where $S_{\mathrm{max}}$ is the maximum von Neumann entanglement entropy generated by the neural network between $A$ and $B$. $\mathcal{E}$ is upper bounded by the Schmidt rank $R$ and, for a given $R$, is maximum for a maximally entangled state. We show that the entangling power of an encoder-decoder neural network depends not only on its latent space dimension (or width) $K$ but also on the decoder complexity. In particular, for polynomial decoders of degree $p$, we show that $\mathcal{E}$ is given by   
\begin{equation}
\mathcal{E}_p(K)=\binom{K+p}{p}.
\label{eq:EpK}
\end{equation}
For a fixed $p$, $\mathcal{E}$ grows polynomially in the latent width $K$, while if $p$ and $K$ grow proportionally, then $\mathcal{E}$ grows exponentially. Eq.~\eqref{eq:EpK} reveals an exponential entangling power of neural networks with modest resources.  

\paragraph*{\textbf{Neural-network encoding}.}
We consider how a general function $f(x,y)$ of two sets of variables $x$ and $y$ can  
be represented as
\begin{equation} 
f(x,y) = g\bigl(\vec\phi_A(x),\vec\phi_B(y)\bigr). 
\label{eq:encoder} 
\end{equation} 
For example, in a many-body wavefunction for spin models, 
$x$ and $y$ can be a bitstring corresponding to spin configurations in spatial region $A$ and $B$ respectively.  
The encoders $\vec\phi_A:x\rightarrow\mathbb C^K$ and $\vec\phi_B:y\rightarrow\mathbb C^K$ process the two input variables separately and map them to vectors in a common $K$-dimensional vector space, which we refer to as the latent space. The decoder $g:\mathbb C^K\times\mathbb C^K\rightarrow\mathbb C$ then combines the two latent vectors $\vec\phi_A(x)$ and $\vec\phi_B(y)$ and maps the concatenated vector to a scalar output. We refer to Eq.~\eqref{eq:encoder} as a bipartite encoder-decoder representation and $K$ as the \textit{latent width}. If $x,y$ themselves belong to $\mathbb{C}^K$, a trivial representation is provided by $\phi_A(x)=x$, $\phi_B(y)=y$ and $g=f$.

A nontrivial encoder-decoder representation is provided by the Schmidt decomposition, which decomposes a function into a sum of products across a bipartition:   
\begin{equation}
f(x,y)=\sum_{\alpha=1}^R \lambda_\alpha u_\alpha(x)v_\alpha(y),
\end{equation}
where $R$ denotes the Schmidt rank. 
This corresponds to the special case of an encoder $x\rightarrow \vec u, y\rightarrow \vec v$ followed by a bilinear decoder with $K=R$: $g(\vec u,\vec v)= \sum_{\alpha=1}^R   \lambda_\alpha u_\alpha v_\alpha$, which is bilinear in the latent variables $u_\alpha(x)$ and $v_\alpha(y)$. 
For $n$ separated Bell pairs, $R=2^n$ grows exponentially with $n$, an example of a highly-entangled state.

A natural question is whether one can obtain a more efficient bipartite representation by allowing the decoder $g(\vec u,\vec v)$ to be a nonlinear function of $\vec u,\vec v$. A nonlinear decoder is common in the context of machine learning, as realized by multilayer feedforward neural networks. 
Can nonlinear decoders enable a bipartite representation of $f(x,y)$ using fewer latent variables than its Schmidt rank, $K<R$? We will show that this is tied to the entangling power of the decoder. 


\par

\paragraph*{\textbf{Maximally entangled example}.} It is instructive to begin with a simple example. Consider $n$ Bell pairs across the bipartition $A|B$, corresponding to the maximally entangled state 
\begin{equation}
\ket{\Psi_{\mathrm{Bell}}} \propto \sum_{x}\ket{x}_A\ket{x}_B,     
\end{equation} 
where each side has $D=2^n$ basis configurations, denoted by the bitstring $x$. 
The corresponding wavefunction is 
\begin{equation} 
f_{\mathrm{Bell}}(x,y) = \frac{\delta_{xy}}{\sqrt D}. \label{eq:bell} \end{equation}
The state has maximum Schmidt rank $D$ and maximum entanglement entropy $S = n \log 2$.
Nevertheless, we show that a single latent variable suffices to write this in the form of Eq.~\eqref{eq:encoder} for sufficiently large degree $p$. 

Let $N(x) \in \{0,\ldots, D-1\} \equiv [D]$ be the integer whose binary representation is $x$, let 
$\omega=e^{2\pi i/D}$, and choose the scalar encoders
\begin{equation}
\phi_A(x)=\omega^{N(x)},
\qquad
\phi_B(y)=\omega^{-N(y)},
\end{equation}
and the decoder
\begin{equation}
g(u,v)
=
D^{-3/2}
\sum_{\ell=0}^{D-1}(uv)^\ell.
\label{eq:root_filter}
\end{equation}
The geometric sum in Eq.~\eqref{eq:root_filter} equals $D$ when $N(x)=N(y)$ or equivalently $x=y$ and vanishes otherwise, so
\begin{equation}
g\bigl(\phi_A(x),\phi_B(y)\bigr)
=
\frac{\delta_{xy}}{\sqrt D}
=
f_{\mathrm{Bell}}(x,y).
\end{equation}
In this construction, the two encoders assign conjugate phase factors according to the basis configurations, while the decoder performs a discrete Fourier sum that tests their equivalence modulo $D$. This encoding using a {\it single} complex latent variable ($K=1$) suffices to distinguish all $D$ possible configurations, i.e., the maps $\phi_{A,B}: [D]\to \mathbb{C}$ are injective. Then, the decoder $g$, a degree-($D-1$) polynomial in $u$ and in $v$ separately, singles out the case $x=y$ uniquely. 
Thus our example serves as a proof of principle that a nonlinear decoder can exactly represent a maximally entangled state with latent width $K=1$. Equivalently, polynomial decoders of degree $D$ (the dimension of Hilbert subspace) have maximal entangling power even for $K=1$.  
\par

Importantly, this extreme compression is not at all special to the maximally entangled state if the decoder is unrestricted. In particular, any injective maps $\phi_{A,B}$ from a discrete set to $\mathbb{C}$ can serve as encoders for any function $f(x,y)$ of discrete variables, provided that the decoder $g$ is chosen to be $f$ composed with the inverse of $\phi_{A,B}$. 
This amounts to embedding $X_{A,B}$ into $\mathbb{C}$ and choosing $g$ to act as a ``lookup table" mapping the image of the embedding to the values of $f$.

\par 
While the maximal compression shown above is always possible if we allow for arbitrary decoder, this generally requires exponentially large complexity in $g$, as shown by Eq.~\eqref{eq:root_filter} for  the maximally entangled state. 

\par 
This motivates us to consider the entangling power of decoders with restricted complexity. In the following, we focus on decoders that are polynomial functions of degree $p$, with $p$ being at most polynomial in $n$. It is then natural to ask: subject to this restriction, what is the entangling power of $g$ at latent width $K$?\par 

Before presenting the general theory, we first present an instructive construction for the Bell state. For $K=n$, we denote the basis labels as bitstrings $x=(x_1\ldots x_n)$ and $y=(y_1\ldots y_n)$, use a string of latent variables $\pm 1$ to encode each bit separately:
\begin{equation}
[\vec\phi_A(x)]_a=(-1)^{x_a},
\qquad
[\vec\phi_B(y)]_a=(-1)^{y_a},
\end{equation}
and define
\begin{equation}
g(\vec u,\vec v)
=
\frac{1}{2^{n/2}}
\prod_{a=1}^n
\frac{1+u_av_a}{2}.
\label{eq:guvprodn}
\end{equation}
Each factor equals one when $x_a=y_a$ and zero otherwise. Hence the product vanishes if any bit differs and equals one only when the two strings are identical, yielding
\begin{equation}
g\bigl(\vec\phi_A(x),\vec\phi_B(y)\bigr)
=
f_{\mathrm{Bell}}(x,y).
\end{equation}
This decoder has degree $p=n$ separately in encoded variables $u_a$ and $v_a$,
 and hence this example teaches us that maximal entangling power is possible with $p=K=\log D$.  
This should be compared with the earlier construction with $p=D$ and $K=1$.   
 
\par

Interestingly, maximal entangling power is possible with a degree $p=\log D$ and a smaller latent width $K_{\mathrm{min}}\simeq 0.29\log D$. Indeed, we now show that one can represent \textit{any} function $f(x,y)$ of discrete variables $x,y \in [D]$ in such a way. 

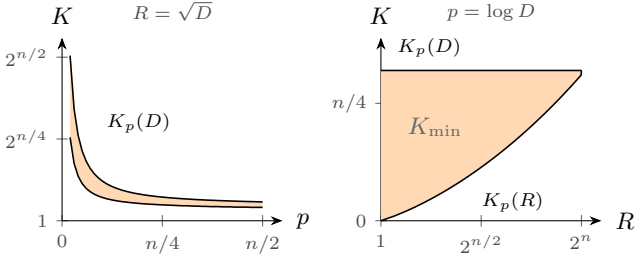
\begin{figure}[t]
\centering
\pgfplotsset{sandwich/.style={
  width=4.5cm, height=4.0cm,
  axis lines=left,
  axis line style={thin,-{Stealth[length=1.6mm]}},
  tick style={thin,black!60},
  ticklabel style={font=\scriptsize,color=black!70},
  label style={font=\small,color=black!70},
  every axis x label/.style={at={(ticklabel* cs:1.0)},anchor=west,xshift=2pt},
  every axis y label/.style={at={(ticklabel* cs:1.0)},anchor=south,yshift=2pt},
  title style={font=\scriptsize,color=black!70,yshift=-3pt},
}}
\tikzset{bound/.style={line width=0.6pt,line join=round}}
\begin{tikzpicture}

\begin{axis}[sandwich, name=axA, clip=false,
  xlabel={$p$}, ylabel={$K$},
  title={$R=\sqrt{D}$},
  xmin=0, xmax=55, xtick={0,25,50}, xticklabels={$0$,$n/4$,$n/2$},
  ymin=0, ymax=56, ytick={0,25,50}, yticklabels={$1$,$2^{n/4}$,$2^{n/2}$},
]
\addplot[draw=none, fill=orange!30] coordinates {%
(2,50.50) (3,34.19) (4,26.15) (5,21.38) (6,18.25) (7,16.04) (8,14.41) (9,13.16) (10,12.18) (11,11.38) (12,10.73) (13,10.19) (14,9.73) (15,9.33) (16,8.99) (17,8.70) (18,8.43) (19,8.20) (20,8.00) (21,7.81) (22,7.65) (23,7.50) (24,7.36) (25,7.23) (26,7.12) (27,7.01) (28,6.92) (29,6.82) (30,6.74) (31,6.66) (32,6.58) (33,6.51) (34,6.44) (35,6.39) (36,6.32) (37,6.27) (38,6.23) (39,6.17) (40,6.13) (41,6.09) (42,6.04) (43,6.00) (44,5.95) (45,5.93) (46,5.88) (47,5.86) (48,5.83) (49,5.78) (50,5.75) (50,4.09) (49,4.09) (48,4.09) (47,4.09) (46,4.17) (45,4.17) (44,4.17) (43,4.17) (42,4.25) (41,4.25) (40,4.25) (39,4.32) (38,4.32) (37,4.32) (36,4.39) (35,4.39) (34,4.46) (33,4.46) (32,4.52) (31,4.58) (30,4.58) (29,4.64) (28,4.70) (27,4.75) (26,4.81) (25,4.86) (24,4.91) (23,5.00) (22,5.04) (21,5.13) (20,5.21) (19,5.32) (18,5.43) (17,5.55) (16,5.70) (15,5.83) (14,6.02) (13,6.23) (12,6.48) (11,6.77) (10,7.13) (9,7.58) (8,8.14) (7,8.89) (6,9.91) (5,11.38) (4,13.65) (3,17.53) (2,25.50)};
\addplot[bound] coordinates {(2,50.50) (3,34.19) (4,26.15) (5,21.38) (6,18.25) (7,16.04) (8,14.41) (9,13.16) (10,12.18) (11,11.38) (12,10.73) (13,10.19) (14,9.73) (15,9.33) (16,8.99) (17,8.70) (18,8.43) (19,8.20) (20,8.00) (21,7.81) (22,7.65) (23,7.50) (24,7.36) (25,7.23) (26,7.12) (27,7.01) (28,6.92) (29,6.82) (30,6.74) (31,6.66) (32,6.58) (33,6.51) (34,6.44) (35,6.39) (36,6.32) (37,6.27) (38,6.23) (39,6.17) (40,6.13) (41,6.09) (42,6.04) (43,6.00) (44,5.95) (45,5.93) (46,5.88) (47,5.86) (48,5.83) (49,5.78) (50,5.75)};
\addplot[bound] coordinates {(2,25.50) (3,17.53) (4,13.65) (5,11.38) (6,9.91) (7,8.89) (8,8.14) (9,7.58) (10,7.13) (11,6.77) (12,6.48) (13,6.23) (14,6.02) (15,5.83) (16,5.70) (17,5.55) (18,5.43) (19,5.32) (20,5.21) (21,5.13) (22,5.04) (23,5.00) (24,4.91) (25,4.86) (26,4.81) (27,4.75) (28,4.70) (29,4.64) (30,4.58) (31,4.58) (32,4.52) (33,4.46) (34,4.46) (35,4.39) (36,4.39) (37,4.32) (38,4.32) (39,4.32) (40,4.25) (41,4.25) (42,4.25) (43,4.17) (44,4.17) (45,4.17) (46,4.17) (47,4.09) (48,4.09) (49,4.09) (50,4.09)};
\node[font=\scriptsize, anchor=west] at (axis cs:9,30) {$K_p(D)$};
\end{axis}

\begin{axis}[sandwich, clip=false, at={($(axA.east)+(1.3cm,0)$)}, anchor=west,
  xlabel={$R$}, ylabel={$K$},
  title={$p=\log D$},
  xmin=0, xmax=110, xtick={0,50,100}, xticklabels={$1$,$2^{n/2}$,$2^{n}$},
  ymin=0, ymax=39, ytick={0,25}, yticklabels={$0$,$n/4$},
]
\addplot[draw=none, fill=orange!30] coordinates {%
(0.00,0) (6.66,1) (12.33,2) (17.43,3) (22.13,4) (26.52,5) (30.67,6) (34.60,7) (38.36,8) (41.96,9) (45.41,10) (48.75,11) (51.97,12) (55.09,13) (58.12,14) (61.06,15) (63.91,16) (66.70,17) (69.41,18) (72.06,19) (74.64,20) (77.17,21) (79.64,22) (82.06,23) (84.43,24) (86.75,25) (89.03,26) (91.26,27) (93.45,28) (95.61,29) (97.72,30) (99.80,31) (100.00,32) (0.00,32)};
\addplot[bound] coordinates {(0,32) (100,32)};
\addplot[bound] coordinates {(0.00,0) (6.66,1) (12.33,2) (17.43,3) (22.13,4) (26.52,5) (30.67,6) (34.60,7) (38.36,8) (41.96,9) (45.41,10) (48.75,11) (51.97,12) (55.09,13) (58.12,14) (61.06,15) (63.91,16) (66.70,17) (69.41,18) (72.06,19) (74.64,20) (77.17,21) (79.64,22) (82.06,23) (84.43,24) (86.75,25) (89.03,26) (91.26,27) (93.45,28) (95.61,29) (97.72,30) (99.80,31) (100.00,32)};
\node[font=\scriptsize, anchor=south west] at (axis cs:4,33) {$K_p(D)$};
\node[font=\scriptsize, anchor=north west] at (axis cs:46,8) {$K_p(R)$};
\node[font=\small, color=black!60] at (axis cs:27,20) {$K_{\min}$};
\end{axis}

\end{tikzpicture}
\caption{Polynomial sandwich bound~\eqref{eq:polynomial_sandwich} for
$D=2^{n}$. (Left) At fixed Schmidt rank $R=\sqrt{D}$, the admissible range of
$K_{\min}$ (shaded) between $K_p(R)$ (lower) and $K_p(D)$ (upper) collapses
as the decoder degree $p$ grows. (Right) At $p=\log D$, the upper bound
$K_p(D)\approx 0.29\,n$ is independent of the Schmidt rank, while the lower
bound $K_p(R)$ grows slowly with the entanglement (continuous interpolation
of integer-valued bounds).}
\label{fig:sandwich}
\end{figure}

\par 
\paragraph*{\textbf{Polynomial decoders}.}
Let $\mathcal P_p$ denote the class of decoders that are polynomials in two vector variables $\vec u$ and $\vec v$ of total degree at most $p$ in each.  
Viewing $f(x,y)$ as a matrix whose rows and columns are indexed by the configurations in $x$ and $y$, respectively, its matrix rank is the Schmidt rank across $A|B$.  For decoders $g\in \mathcal P_p$, we now prove the following theorem that relates this rank to the minimum latent width.

\medskip
\noindent\textbf{Theorem.}
Consider representations
\begin{equation}
f(x,y)
=
g\bigl(\vec\phi_A(x),\vec\phi_B(y)\bigr),
\quad
x\in X_A,\,\, y\in X_B,
\end{equation}
where $X_A$ and $X_B$ are finite sets having $D_A$ and $D_B$ elements, respectively, the encoders $\vec\phi_A:X_A\to\mathbb C^K$ and $\vec\phi_B:X_B\to\mathbb C^K$ are arbitrary, and the decoder $g\in\mathcal P_p$. Then the entangling power of degree-$p$ decoder class is $\mathcal E_p(K) = \mathcal{N}_p(K)$ with \begin{equation}
\mathcal N_p(K) \equiv 
\binom{K+p}{p}
\label{eq:monomial_count}
\end{equation}
the integer counting the number of independent monomials in $K$ variables of total degree at most $p$, provided $D_A,D_B \geq \mathcal{N}_p(K)$. 

Consequently, letting $K_{\min}$ be the smallest latent width for which such a representation exists, $R$ be the matrix rank of $f(x,y)$ and $D=\max\{D_A, D_B \}$, we have
\begin{equation}
K_p(R)
\leq
K_{\min}
\leq
K_p(D),
\label{eq:polynomial_sandwich}
\end{equation}
where $K_p(M) \equiv \min\bigl\{
K\in\mathbb N:
\mathcal N_p(K)\geq M
\bigr\}$. This is illustrated in Fig.~\ref{fig:sandwich}. In terms of the entangling power, this is equivalent to
\begin{equation}
    R\leq \mathcal{E}_p(K_{\mathrm{min}}),\quad\mathcal{E}_p(K_{\mathrm{min}}-1) \leq D.
\label{eq:polynomial_sandwich2}
\end{equation}

The lower bounds in Eqs.~\eqref{eq:polynomial_sandwich} and \eqref{eq:polynomial_sandwich2} establish that the entangling power is a key quantity for neural networks. 
On the one hand, it is necessary for it to exceed the Schmidt rank of the target function $f(x,y)$. On the other hand, any $f$ can be represented once the entangling power is large enough, which we show by explicit construction.

\medskip
We will prove Eq.~\eqref{eq:polynomial_sandwich}, from which Eq.~\eqref{eq:polynomial_sandwich2} follows. We first prove the lower bound. Let $m_1,\ldots,m_{\mathcal N}$, with $\mathcal N=\mathcal N_p(K)$, denote the monomials in $K$ variables of total degree at most $p$. Every decoder $g\in\mathcal P_p$ can be expanded as
\begin{equation}
g(\vec u,\vec v)
=
\sum_{a,b=1}^{\mathcal N}
C_{ab}m_a(\vec u)m_b(\vec v)
=
\sum_{a=1}^{\mathcal N}
m_a(\vec u)h_a(\vec v),
\label{eq:poly_product_expansion}
\end{equation}
where $h_a(\vec v)=\sum_b C_{ab}m_b(\vec v)$. Combining it with the encoders, the bipartite  representation of $f$ takes the form
\begin{equation}
f(x,y)
=
\sum_{a=1}^{\mathcal N}
m_a\bigl(\vec\phi_A(x)\bigr)
h_a\bigl(\vec\phi_B(y)\bigr).
\label{eq:poly_product_functions}
\end{equation}
This is a sum of at most $\mathcal N_p(K)$ product functions across $A|B$, so the Schmidt rank of $f$ cannot exceed $\mathcal N_p(K)$. Representing a target function of Schmidt rank $R$ therefore requires $R\leq\mathcal N_p(K)$, or equivalently $K_p(R)\leq K$. This proves the lower bound.

We next prove the upper bound. We seek an encoder-decoder construction satisfying
\begin{equation}
g\bigl(\vec\phi_A(x),\vec\phi_B(y)\bigr)
=
f(x,y)
\qquad
\text{for all }x\in X_A,\ y\in X_B.
\label{eq:full_interpolation_problem}
\end{equation}
Assuming that there exists a set of degree$-p$ polynomials indexed by $x$, $q_x^{(A)}:\mathbb{C}^K\to \mathbb{C}$, and likewise, there exists a set of  degree$-p$ polynomials indexed by $y$, $q_y^{(B)}:\mathbb{C}^K\to \mathbb{C}$,  satisfying
\begin{equation}
q_x^{(A)}
\bigl(\vec\phi_A(x')\bigr)
=
\delta_{xx'},
\qquad
q_y^{(B)}
\bigl(\vec\phi_B(y')\bigr)
=
\delta_{yy'}, 
\label{eq:indicator}
\end{equation}
then it is clear that we can choose the decoder as 
\begin{equation}
g(\vec u,\vec v)
=
\sum_{\substack{x\in X_A\\y\in X_B}}
f(x,y)\,
q_x^{(A)}(\vec u)
q_y^{(B)}(\vec v).
\label{eq:indicator_decoder}
\end{equation}
This decoder belongs to $\mathcal P_p$, since it has degree at most $p$ separately in $\vec u$ and $\vec v$. Evaluating it on $\vec\phi_A(x'),\vec \phi_B(y')$ gives $f(x',y')$. Thus, it remains only to construct such polynomials $q_x^{(A)},q_y^{(B)}$, which we refer to as \textit{indicator} polynomials.
\par

We now prove that the desired set of indicator polynomials indeed exists when $\mathcal N\geq D$. 
Intuitively, a degree-$p$ polynomial $q^{(A)}(\vec u)$ has $\mathcal N$ adjustable coefficients. Prescribing its output values on $D$ input values $\vec u_x \equiv \phi_A(x)$ for all $x \in X_A$ gives $D$ linear conditions on the coefficients. This linear system can be solved when the number of unknowns $\mathcal N$ is greater than the number of equations $D$.    

To prove it rigorously, we expand $q^{(A)}(\vec u)$ in the monomial basis
\begin{equation}
q^{(A)}(\vec u)
=
\sum_{a=1}^{\mathcal N}
c_a m_a(\vec u).
\end{equation}
Given arbitrary target values $\vec t\in\mathbb C^D$, 
the interpolation conditions
\begin{equation}
q^{(A)}\bigl(\vec u_{x'}\bigr)
=
t_{x'},
\qquad
x'\in X_A,
\label{eq:one_side_interpolation}
\end{equation}
are equivalent to the linear system
\begin{equation}
M\vec c
=
\vec t,
\qquad
[M]_{x'a}
=
m_a\bigl(\vec u_{x'}\bigr). 
\label{eq:evaluation_matrix}
\end{equation}
Here, $M$ is a $D\times\mathcal N$ matrix indexed by $x'$ and $a$, which tabulates the values of all $\mathcal N$ monomials evaluated at $D$ latent vectors $\vec u_{x'}$ indexed by $x'\in X_A$.   

Provided that $M$ has full row rank $D$,  the linear system \eqref{eq:evaluation_matrix} can be solved for any $\vec t\in\mathbb C^D$. In particular, choosing $\vec t=\vec e_x$, the vector that is $1$ on configuration $x$ and $0$ on all others, gives the desired indicator polynomial $
q_x^{(A)}
\bigl(\vec u_{x'}\bigr)
=
\delta_{xx'}.$ The same argument applies on side $B$. Then, we can construct degree-$p$ decoder for any target function. 

The final step is to show that when $\mathcal N \geq D$ is satisfied, one can indeed find such matrix $M$ with full row rank $D$.  To see this, consider the evaluation map $\mathbb C^{K} \rightarrow \mathbb C^\mathcal{N}$: 
\begin{equation}
\vec m(\vec u) \equiv (m_1(\vec u), ..., m_{\mathcal N}(\vec u)), \label{evaluation}
\end{equation}
which maps an input vector $\vec u$ to an $\mathcal N$-component vector $\vec m$ that lists the values of monomials evaluated at $\vec u$. Importantly, the full set of $\vec m$ generated from all points $\vec u$ in the latent space {\it spans} the vector space $\mathbb C^{\mathcal N}$. 
This allows us to choose $\mathcal N$ latent vectors as the image of the encoder, such that these $\mathcal N$ points are mapped by Eq.~\eqref{evaluation} to $\mathcal N$ linearly independent vectors in $\mathbb{C}^{\mathcal N}$. 
Consequently, when $\mathcal N\geq D$, we can choose a subset of these $D$ points that are mapped to $D$ linearly independent vectors in $\mathbb{C}^{\mathcal N}$. Labeling these $D$ latent vectors $\vec u_{x'}$ with $x'\in X_A$ and stacking them together as rows gives the desired $D \times \mathcal N$ matrix $M$, which has full row rank. 

We have thus proven that any function $f(x,y)$ can be represented exactly using a decoder that is a polynomial of total degree up to $p$, provided that $\mathcal N \geq D$ is satisfied. By definition, $K_p$ is the smallest integer satisfying $\mathcal N_{p}(K_p) \geq D$. Hence $K_{\min}\le K_p(D)$.

The upper and lower bounds need not coincide in general, but do for maximally entangled states, for which $R=D$. In particular, for the Bell state we have $R=D=2^n$. The minimum latent-space dimension is therefore determined exactly by $K_{\min}=K_p(2^n)$. Three representative choices of the decoder degree give
\begin{equation}
K_{\min}=
\begin{cases}
2^n-1, & p=1,\\[3pt]
0.2938\,n, & p=n \textrm{ and } n\gg 1,\\[3pt]
1, & p\geq 2^n-1.
\end{cases}
\label{eq:bell_polynomial_profile}
\end{equation}
Importantly, these three regimes make the tradeoff between polynomial degree and latent width explicit: a linear decoder requires an exponentially large latent space, a degree-$n$ decoder requires only a linear-in-$n$ latent space, and an exponentially large degree reduces the latent space to a single variable. More generally, between the first two regimes, the width scales as
\begin{equation}
K_{\min}\sim (p!)^{1/p}2^{n/p},
\label{eq:bell_intermediate_profile}
\end{equation}
which becomes $K_{\min}\sim (p/e)2^{n/p}$ when $1\ll p\ll n$. Thus, increasing the decoder degree continuously reduces the exponential cost before the width becomes linear at $p=n$.
The asymptotic estimates and the factor $0.2938$ are derived in the Supplemental Material.

\paragraph*{\textbf{Neural-network implementation}.} Thus far, our results have been obtained for general polynomial decoders, independent of the specific implementation of the neural architecture. The factored polynomial in Eq.~\eqref{eq:guvprodn} admits a compact neural-network implementation. Each factor $r_a=(1+u_av_a)/2$ tests whether the two bit strings agree at position $a$. With the quadratic activation $\sigma(z)=z^2$, multiplication can be implemented exactly as
\begin{equation}
\operatorname{Mult}(u,v)
=
\frac{\sigma(u+v)-\sigma(u-v)}{4}
=
uv.
\label{eq:quadratic_multiplication}
\end{equation}
A balanced binary tree of these modules therefore evaluates Eq.~\eqref{eq:guvprodn} using $O(n)$ units and depth $O(\log n)$.


More generally, 
a maximally entangled state takes the form 
\begin{equation}
f_U(x,y)=U_{yx}/\sqrt{D}, \label{max}
\end{equation}
where $U$ is a unitary matrix indexed by $x,y$. 
Eq.~\eqref{max} admits a compact polynomial decoder whenever the function $(x,y)\to U_{yx}$ has a polynomial-size arithmetic-circuit representation. This includes structured examples such as $U=I$, 
but not a generic maximally entangled state.

In conclusion, we have introduced \textit{entangling power} as a complexity metric for encoder-decoder neural networks. The entangling power quantifies the capability of such neural networks to reconstruct a quantum wavefunction, or more general function, from two partitioned parts. We have shown that the nonlinearity of the decoder empowers neural networks to capture highly-entangled states with exponentially fewer resources.  

\begin{acknowledgments}
\textit{Acknowledgments}. This work was supported by a Simons Investigator Award from the Simons Foundation. T.W. is grateful for the support by the Harvard Quantum
Initiative Fellowship and the Simons Collaboration on Ultra-Quantum Matter, which is a grant from the Simons Foundation (Grant No. 651440). NP acknowledges support
from the Walter Burke Institute for Theoretical Physics and Institute for Quantum Information and Matter, an NSF Physics Frontiers Center (PHY-2317110).
\end{acknowledgments}

\bibliographystyle{apsrev4-2}
\bibliography{short}

\clearpage
\onecolumngrid

\begin{center}
{\large\bf Supplemental Material:\\[2mm]
Entangling power of neural networks}

\end{center}
\vspace{2cm}

Here, we present the theorem which was discussed in the main text in self-contained and complete form, along with corollaries. Throughout, the encoders $\vec\phi_A:X_A\to\mathbb C^K$ and $\vec\phi_B:X_B\to\mathbb C^K$ are arbitrary and the decoder lies in $\mathcal P_p$, the class of polynomials of total degree at most $p$ separately in $\vec u$ and in $\vec v$. We write $D_A=|X_A|$, $D_B=|X_B|$, $D=\max\{D_A,D_B\}$, and let $R$ denote the rank of the matrix $[f(x,y)]$, i.e.\ the Schmidt rank across $A|B$. Recall
\begin{equation}
\mathcal N_p(K)=\binom{K+p}{p},
\qquad
K_p(M)=\min\{K\in\mathbb N:\mathcal N_p(K)\geq M\}.
\end{equation}

\begin{theorem}[Theorem of the main text]
\label{prop:polydecoder}
An exact representation $f(x,y)=g(\vec\phi_A(x),\vec\phi_B(y))$ with $K$ latent variables and $g\in\mathcal P_p$ requires
\begin{equation}
R\leq\mathcal N_p(K).
\label{eq:poly_necessary_SM}
\end{equation}
Conversely, if $\mathcal N_p(K)\geq D$, then every $f:X_A\times X_B\to\mathbb C$ admits such a representation. Consequently
\begin{equation}
K_p(R)\leq K_{\min}\leq K_p(D).
\label{eq:poly_bounds_SM}
\end{equation}
\end{theorem}

\begin{proof}
Let $m_1,\ldots,m_{\mathcal N}$, with $\mathcal N=\mathcal N_p(K)$, be the monomials in $K$ variables of total degree at most $p$. Expanding $g$ in its first argument gives $g(\vec u,\vec v)=\sum_{a=1}^{\mathcal N}m_a(\vec u)h_a(\vec v)$. Inserting the encoders then expresses $f$ as a sum of at most $\mathcal N$ product functions across $A|B$, which proves Eq.~\eqref{eq:poly_necessary_SM}.
 
For the converse, consider the evaluation map $\vec m(\vec u)=(m_1(\vec u),\ldots,m_{\mathcal N}(\vec u))$ from $\mathbb C^K$ to $\mathbb C^{\mathcal N}$. Its image spans $\mathbb C^{\mathcal N}$. Indeed, otherwise there would exist a nonzero $\vec c\in\mathbb C^{\mathcal N}$ with $\sum_a c_a m_a(\vec u)=0$ for all $\vec u$, i.e.\ a nonzero polynomial vanishing identically on $\mathbb C^K$, which is impossible. Since $\mathcal N\geq D\geq D_A$, we may therefore choose points $\{\vec a_x\}_{x\in X_A}$ in $\mathbb C^K$ whose images $\vec m(\vec a_x)$ are linearly independent. The $D_A\times\mathcal N$ matrix
\begin{equation}
[V_A]_{xa}=m_a(\vec a_x)
\end{equation}
then has full row rank, and hence a right inverse $Q_A$ satisfying $V_AQ_A=I$. The same construction on side $B$ gives points $\{\vec b_y\}_{y\in X_B}$, a matrix $V_B$, and a right inverse $Q_B$. Define the indicator polynomials
\begin{equation}
q^{(A)}_x(\vec u)=\sum_{a=1}^{\mathcal N}[Q_A]_{ax}m_a(\vec u),
\qquad
q^{(B)}_y(\vec v)=\sum_{a=1}^{\mathcal N}[Q_B]_{ay}m_a(\vec v),
\end{equation}
which satisfy $q^{(A)}_x(\vec a_{x'})=\delta_{xx'}$ and $q^{(B)}_y(\vec b_{y'})=\delta_{yy'}$. Choosing $\vec\phi_A(x)=\vec a_x$, $\vec\phi_B(y)=\vec b_y$, and
\begin{equation}
g(\vec u,\vec v)=\sum_{x\in X_A}\sum_{y\in X_B}f(x,y)\,q^{(A)}_x(\vec u)\,q^{(B)}_y(\vec v)\in\mathcal P_p
\end{equation}
which, it is straightforward to check, yields $g(\vec\phi_A(x),\vec\phi_B(y))=f(x,y)$. Eq.~\eqref{eq:poly_bounds_SM} follows by minimizing over $K$.
\end{proof}

\begin{corollary}[Entangling power]
\label{prop:entangling_power}
Let $D_A,D_B\geq\mathcal N_p(K)$. The maximum entanglement entropy across $A|B$ of a normalized state representable with $K$ latent variables and $g\in\mathcal P_p$ is $S_{\max}=\log\mathcal N_p(K)$, and hence
\begin{equation}
\mathcal E_p(K)=e^{S_{\max}}=\mathcal N_p(K).
\end{equation}
\end{corollary}
 
\begin{proof}
For a state of Schmidt rank $R$ we have $S\leq\log R$, with equality only for a flat entanglement spectrum, and $R\leq\mathcal N_p(K)$ by Eq.~\eqref{eq:poly_necessary_SM}. Hence $S\leq\log\mathcal N_p(K)$. To see that the bound is attained, restrict to subsets of $X_A$ and $X_B$ of size $\mathcal N=\mathcal N_p(K)$ and take $f(x,y)=U_{yx}/\sqrt{\mathcal N}$ with $U$ unitary. This state has $R=\mathcal N$ and a flat spectrum, and by the converse of Thm.~\ref{prop:polydecoder} with $D=\mathcal N$ it admits a representation with $K$ latent variables and $g\in\mathcal P_p$.
\end{proof}

\begin{corollary}[Bell state]
\label{cor:bell_polynomial_SM}
For $n$ Bell pairs, $R=D=2^n$, the two bounds in Eq.~\eqref{eq:poly_bounds_SM} coincide, and
\begin{equation}
K_{\min}=K_p(2^n)=\min\left\{K:\binom{K+p}{p}\geq2^n\right\}.
\end{equation}
\end{corollary}

We record the three regimes quoted in the main text. For fixed $p$,
\begin{equation}
\mathcal N_p(K)=\frac{K^p}{p!}\left(1+O(K^{-1})\right),
\end{equation}
so that $K_{\min}=(p!)^{1/p}2^{n/p}(1+o(1))$. For $1\ll p\ll n$ this becomes $K_{\min}\sim(p/e)\,2^{n/p}$, using $(p!)^{1/p}=(p/e)(1+o(1))$. For $p=n$ and $K=\alpha n$, Stirling's formula gives
\begin{equation}
\frac1n\log\mathcal N_n(\alpha n)=(1+\alpha)\log(1+\alpha)-\alpha\log\alpha+o(1),
\end{equation}
so the threshold $\mathcal N_n(K)\geq2^n$ approaches $\alpha=\alpha_0$, the unique root of
\begin{equation}
(1+\alpha)\log(1+\alpha)-\alpha\log\alpha=1,
\qquad
\alpha_0=0.293815\ldots
\label{eq:alpha0}
\end{equation}
Finally $\mathcal N_p(1)=p+1$, so a single latent variable suffices exactly when $p\geq2^n-1$. At $p=1$ we have $\mathcal N_1(K)=K+1$ and hence $K_{\min}=2^n-1$, one less than the bilinear value $K_{\min}=R=2^n$. The difference originates from the constant term, which is present in $\mathcal P_1$ but absent from a bilinear decoder.

\end{document}